\documentclass[prb,twocolumn,superscriptaddress]{revtex4}

\usepackage{amsmath}

\usepackage{graphicx}
\usepackage{times}

\begin{document}

\date{today}

\title{An exactly soluble model with tunable $p$-wave paired fermion ground states}

\author{Yue Yu}
\affiliation{Institute of Theoretical Physics, Chinese Academy of
Sciences, P.O. Box 2735, Beijing 100080, China}
\author{Ziqiang Wang}
\affiliation{Department of Physics, Boston College, Chestnut Hill,
MA 02467}
\date{\today}
\begin{abstract}
Motivated by the work of Kitaev, we construct an exactly soluble
spin-$\frac{1}2$ model on honeycomb lattice whose ground states
are identical to
$\Delta_{1x}p_x+\Delta_{1y}p_y+i(\Delta_{2x}p_x+\Delta_{2y}p_y)$-wave
paired fermions on square lattice, with tunable paring order
parameters. We derive a universal phase diagram for this general
$p$-wave theory which contains a gapped A phase and a
topologically non-trivial B phase. We show that the gapless
condition in the B phase is governed by a generalized inversion
(G-inversion) symmetry under $p_x\leftrightarrow {\Delta_{1y}\over
\Delta_{1x}} p_y$.  The G-inversion symmetric gapless B phase near
the phase boundaries is described by 1+1-dimensional gapless
Majorana fermions in the asymptotic long wave length limit, i.e.
the $c=1/2$ conformal field theory. The gapped B phase has
G-inversion symmetry breaking and is the weak pairing phase
described by the Moore-Read Pfaffian. We show that in the gapped B
phase, vortex pair excitations are separated from the ground state
by a finite energy gap.
\end{abstract}


\maketitle

\section{Introduction}

The low energy excitations of a topologically non-trivial phase have
remarkable properties. A well-known example is the quasihole
excitation of the Laughlin state in the fractional quantum Hall
effect (FQHE), which carries fractional charge and anyon statistics.
The most intriguing possibility of a topological phase of matter is
the nonabelian FQHE proposed by Moore and Read \cite{mr} for
even-denominator filling factors, e.g. $\nu=\frac{5}2$ \cite{xia}.
The quasiparticle excitations, vortices in the Moore-Read Pfaffian
wave function, have non-abelian statistics \cite{mr} which plays a
fundamental role in topological quantum computation \cite{ki1,ki2}.
The key ingredient in the Moore-Read Pfaffian state is that the
topologically nontrivial part of the wave function is
asymptomatically the same as the pair wave function in a
$p_x+ip_y$-wave fermion paired state \cite{GWW} in the weak pairing
phase \cite{rg}. The existence of the exotic non-abelian statistics
is thus likely a generic property of the more tangible time-reversal
symmetry (T-symmetry) breaking p-wave pairing states.

Recently, Kitaev constructed a spin-$\frac{1}2$ model with
link-dependent Ising couplings on the honeycomb lattice
\cite{ki2}. Kitaev showed that the model is equivalent to a
bilinear Majorana fermion model and is thus exactly soluble. A
topological non-trivial gapless phase (the B phase) was
discovered.(A Jordan-Wigner transformation to a model with
two-Majorana fermions for this model has been proposed in
\cite{fzx}). In the presence of a T-symmetry breaking term, the B
phase becomes gapped and exhibits vortex excitations obeying
nonabelian statistics. The model also has a topologically trivial,
gapped A phase. The two phases are separated by a topological
phase transition via a gapless critical state. These properties
strongly resemble the weak and strong pairing phases and the
critical state in the $p_x+ip_y$-wave paired states of spinless
fermions \cite{rg}.

The interconnections among the Kitaev model, the $p$-wave paired
fermions, and the Moore-Read Pfaffian and its excitations have not
been well understood previously. In particular, it is important to
understand the universal properties among these systems, analogous
to finding the universality class in statistical mechanics models.
In this paper, we show that the Kitaev model is a special case of
a broader class of two-dimensional spin-$\frac{1}2$ models whose
ground states are equivalent to general paired fermion states in
the p-wave channel. Indeed, the vortex-free Kitaev Hamiltonian
maps to an exact BCS fermion pairing model with $i(p_x+p_y)$-wave
attractions on a square lattice \cite{cn}. Our generalized model
includes both the $p_x+ip_y$ wave paired states and the original
Kitaev model as special limits. It is an exactly soluble model
with minimal three and four-spin interactions. We show that the
vortex-free ground states of this model are described by
$\Delta_{1x}p_x+\Delta_{1y}p_y+i(\Delta_{2x}p_x+\Delta_{2y}p_y)$-wave
paired fermion states with tunable pairing order parameters
$\Delta_{ab}$ on a square lattice. We find that the structure of
the phase diagram is determined by the geometry of the underlying
Fermi surface. It contains both topologically trivial (A) and
nontrivial (B) phases. The A phase is always gapped and
corresponds to the strong pairing phase. The B phase can be either
gapped or gapless even if T-symmetry is broken. We find that
gapless excitations in the B phase is protected by a generalized
inversion (G-inversion) symmetry under $p_x\leftrightarrow
{\Delta_{1y}\over \Delta_{1x}} p_y$ and the emergence of a gapped
B phase is thus tied to G-inversion symmetry breaking. For
instance, the $p_x+ip_y$ wave paired state is gapped while
$p_y+ip_y$-wave paired state is gapless although they both break
the T-symmetry. The critical states of the A-B phase transition
remains gapless whether or not T- and G-inversion symmetries are
broken, indicative of its topological nature. Indeed, if all
$\Delta_{ab}$ are tuned to zero, the topological A-B phase
transition is from a band insulator to a free Fermi gas. The Fermi
surface shrinks to a point zero at criticality.

We show that the gapped B phase is a weak pairing state while the
G-inversion symmetric ground states are extended. The gapless
phase was not well-understood before. We show that the effective
theory near the phase boundary corresponds to 1+1-dimensional
massless Majorana fermions in the long wave length limit, i.e., a
$c=1/2$ conformal field theory or the 2-dimensional Ising model.
The vortex excitations are important in the family of Kitaev
models since the vortex excitations may obey anyon statistics
\cite{ki2}. The vortex excitation energies have been numerically
estimated in the A phase \cite{vidal} and the B phase
\cite{pachos}. We study the vortex excitations in the gapped B
phase in the continuum limit and show that the vortex pair
excitations cost a finite energy. This is consistent with the
results of numerical calculations \cite{pachos} and suggests that
vortex excitations may have well-defined statistics.

\section{The generalization of Kitaev model}

\begin{figure}[htb]
\begin{center}
\includegraphics[width=4cm]{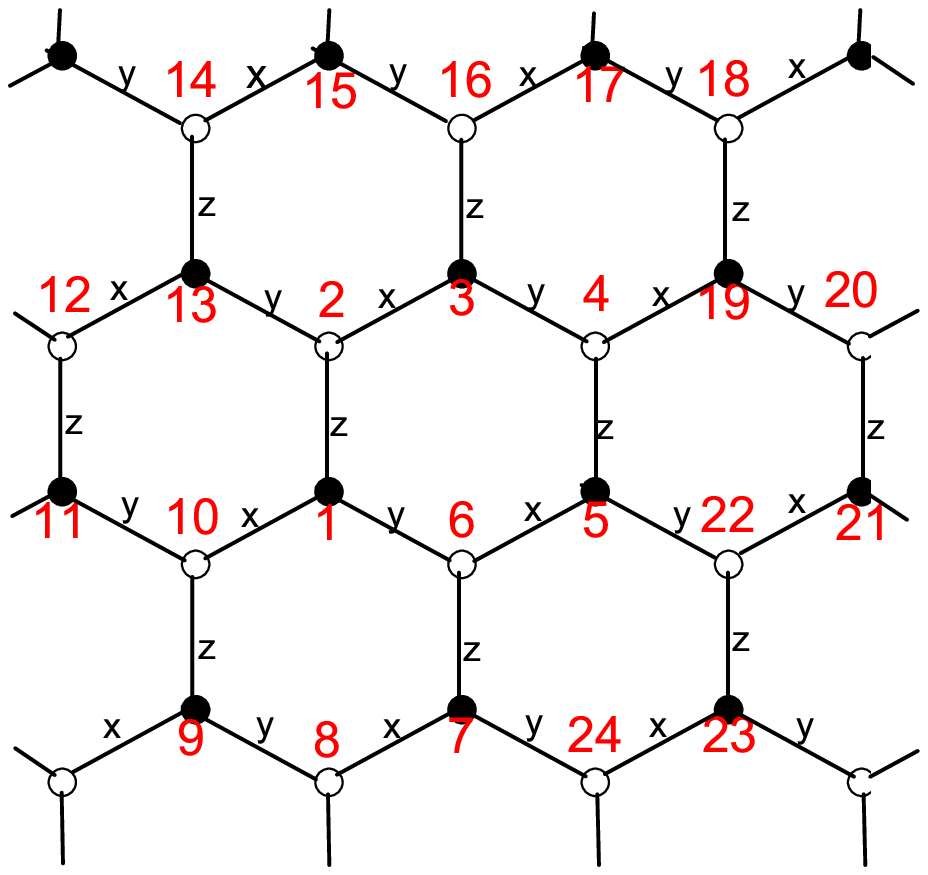}
\includegraphics[width=5cm]{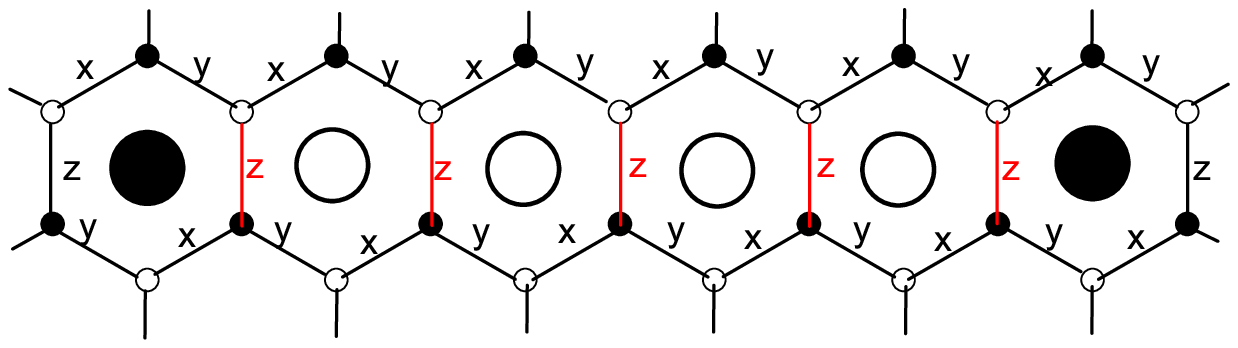}
\end{center}
\vspace{-0.5cm}
 \caption{\label{fig:Fig. 1} (Color online) Upper panel: The honeycomb lattices and links .
 Lower panel: The vortex
excitations. The empty circles denote $W_P=1$ and the filled
circles denote vortices with $W_P=-1$.}
\end{figure}

We extend the Kitaev model on the honeycomb lattice by introducing
minimal three- and four-spin terms in the Hamiltonian,
\begin{eqnarray}
H&=&-J_x\sum_{x-links}\sigma_i^x
\sigma_j^x-J_y\sum_{y-links}\sigma_i^y
\sigma_j^y-J_z\sum_{z-links}\sigma_i^z \sigma_j^z\nonumber\\
&-&\kappa_x\sum_b
\sigma^z_b\sigma^y_{b+e_z}\sigma^x_{b+e_z+e_x}\nonumber\\&
-&\kappa_x\sum_w
\sigma^x_{w}\sigma^y_{w+e_x}\sigma^z_{w+e_x+e_z}\nonumber\\&-&\kappa_y\sum_b
\sigma^z_{b}\sigma^x_{b+e_z}\sigma^y_{b+e_z+e_y}\nonumber\\
&-&\kappa_y\sum_w
\sigma^y_{w}\sigma^x_{w+e_y}\sigma^z_{w+e_y+e_z}\nonumber\\
&-&\lambda_x\sum_{b}\sigma_{b}^z\sigma^y_{b+e_z}
\sigma_{b+e_z+e_x}^y\sigma^z_{b+e_z+e_x+e_z}\nonumber\\&-&\lambda_y\sum_{b}
\sigma_{b}^z\sigma^x_{b+e_z}
\sigma_{b+e_z+e_y}^x\sigma^z_{b+e_z+e_y+e_z},\label{gk}
\end{eqnarray}
where  $\sigma^{x,y,z}$ are Pauli matrices, $x$-,$y$-,$z$-links
are shown in Fig.~\ref{fig:Fig. 1}(upper panel), $'w'$ and $'b'$
label the white and black sites of lattice, and $e_x,e_y,e_z$ are
the positive unit vectors, which are defined as, e.g.,
$e_{12}=e_z,e_{23}=e_x,e_{61}=e_y$. $J_{x,y,z}$, $\kappa_{x,y}$
and $\lambda_{x,y}$ are tunable real parameters. The original
Kitaev model has $\kappa_\alpha=\lambda_\alpha=0$, $\alpha=x,y$.
Adding a T-symmetry breaking external magnetic field corresponds
to $\kappa_x=\kappa_y=\kappa\ne0$ and a $\kappa_z$-term
\cite{ki2,fzx}. It is important to note that the generalized
Hamiltonian maintains the $Z_2$ gauge symmetry acted by a group
element, e.g.,
$$W_P=\sigma^x_1\sigma^y_2\sigma^z_3\sigma^x_4\sigma^y_5\sigma^z_6$$
with $[H,W_P]=0$. In fact, one can construct $Z_2$ gauge invariant
spin models with higher multi-spin terms, e.g.,
$\sigma_{9}^z\sigma_{10}^y\sigma^y_1\sigma^y_2\sigma^x_3,
\sigma_{9}^z\sigma_{10}^y\sigma^y_1\sigma^y_2\sigma^z_3\sigma^y_4$
and $
\sigma_{9}^z\sigma_{10}^y\sigma^y_1\sigma^y_2\sigma^y_3\sigma^z_{16}$,
and so on. One can also add the 'z'-partners of $\kappa_{x,y}$ and
$\lambda_{x,y}$ terms so that the model becomes more symmetric.
However, adding these term or not will not affect our result in
this paper as we will explain later.

We now write down the Majorana fermion representation of this spin
model. Let $b_{x,y,z}$ and $c$ be the four kinds of Majorana
fermions with $b_{x,y,z}^2=1$ and $c^2=1$. The spin operator is
given by $$ {\hat
\sigma^a}=\frac{i}2(b_ac-\frac{1}2\epsilon_{abc}b_bb_c). $$
Restricting to the physical Hilbert space, one needs to require
\cite{ki2}
$$D=b_xb_yb_zc=1$$  and thus $\hat\sigma^a=ib_a c$. The
Hamiltonian now reads
\begin{eqnarray}
H &=&i\sum_a \sum_{a-links} J_au^a_{ij} c_ic_j
-i\sum_b K^x_{b,b+e_z}c_bc_{b+e_z+e_x}\nonumber\\
&-&i\sum_wK^x_{w+e_x-e_x,w-e_z-e_x}c_{w+e_x-e_x}c_{w-e_z-e_x}\nonumber\\
&-&i\sum_{b}\Lambda^x_{b,b+2e_z+e_x}c_{b}c_{b+2e_z+e_x}
\nonumber\\&-&i\sum_{w}\Lambda^x_{w,w-2e_z-e_x}c_{w}c_{w-2e_z-e_x}
\nonumber\\&+&y{\rm -partners}
\end{eqnarray}
where $K^x_{b,b+e_z}=\kappa_x
u^z_{b,b+e_z}u^x_{b+e_z+e_x,b+e_z},~\Lambda^x_{b,b+2e_z+e_x}=\lambda_x
u^z_{b,b+e_z}u^x_{b+e_z,b+e_z+e_x}u^z_{b+e_z+e_x,b+e_z+e_x+e_z}$
etc and $u^a_{ij}=ib_i^ab_j^a$ on $a$-links. It can be shown that
the Hamiltonian commutes with $u^a_{ij}$ and thus the eigenvalues
of $u^a_{ij}=\pm 1$ because $(u_{ij}^a)^2=1$. Since the four spin
terms we introduced are related to the hopping between the 'b' and
'w' sites, Lieb's theorem \cite{lieb} is still applicable. The
third spin terms are 'b' to 'b' and 'w' to 'w' and Lieb's theorem
is not directly applicable. However, according to Kitaev
\cite{ki2}, one can still take $u^a_{bw}=-u^a_{wb}=1$. The
Hamiltonian for the ground state free of the $Z_2$ vortices
($W_P=1$ for all $P$) is given by
\begin{eqnarray}
H_0 &=&i\tilde
J_x\sum_s(c_{s,b}c_{s-e_x,w}-c_{s,w}c_{s-e_x,b})\nonumber\\&+&i\tilde
\lambda_x\sum_s(c_{s,b}c_{s-e_x,w}+c_{s,w}c_{s-e_x,b}) \nonumber\\
&+&i\frac{\kappa^x}2\sum_s
(c_{s,b}c_{s+e_x,b}+c_{s,w}c_{s-e_x,w})\nonumber\\
&+&y~{\rm partners}+ iJ_z\sum_s c_{s,b}c_{s,w} \label{gh}
\end{eqnarray}
where $s$ represents the position of a $z$-link, $\tilde
\lambda_\alpha=\frac{J_\alpha+\lambda_\alpha}2$ and $\tilde
J_\alpha=\frac{J_\alpha-\lambda_\alpha}2$.

\begin{figure}[htb]
\begin{center}
\vspace{-0.5cm}
\includegraphics[width=4cm]{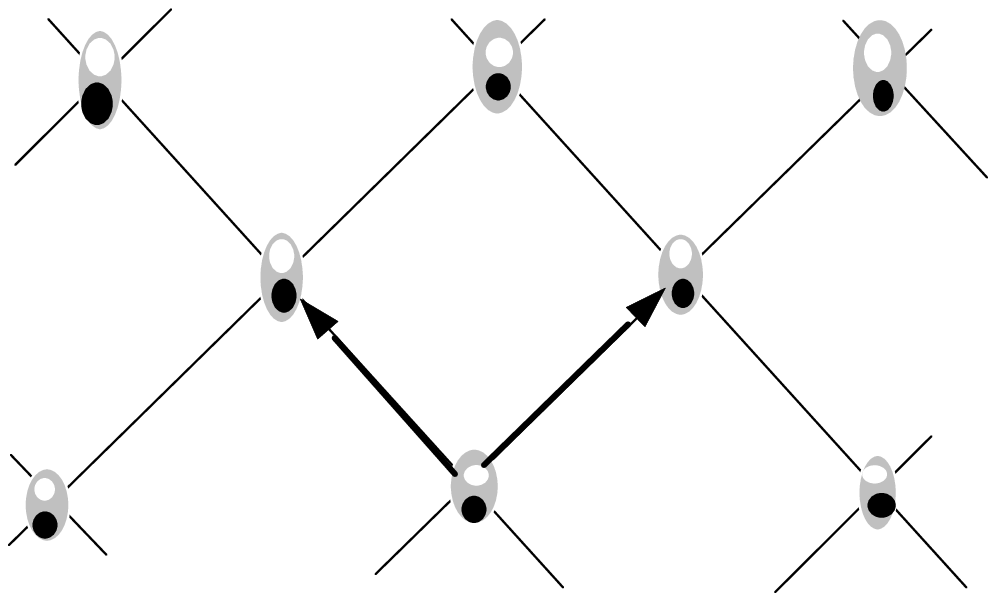}
\includegraphics[width=4cm]{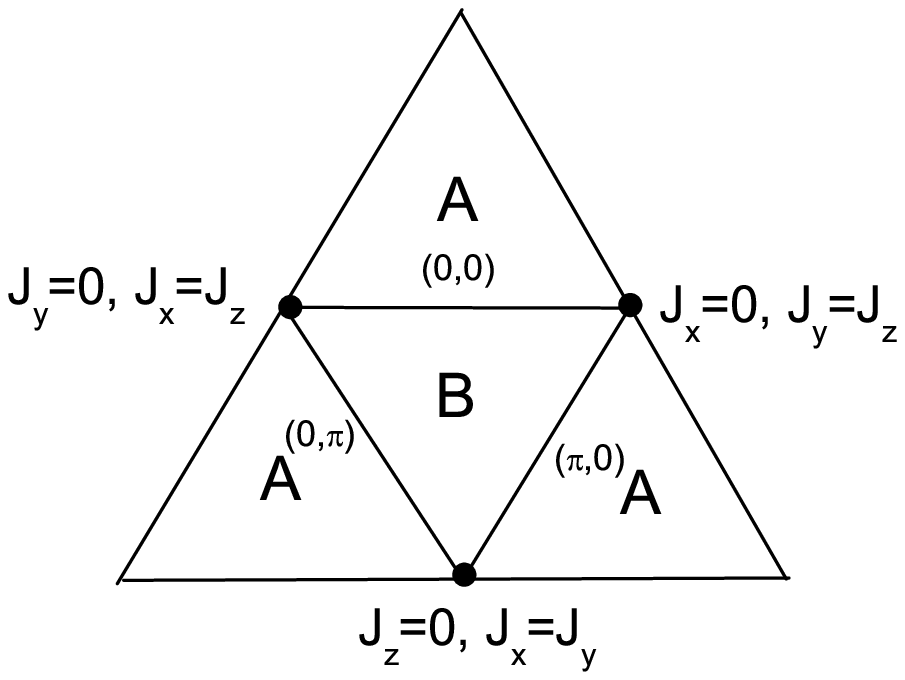}
\end{center}
 \caption{\label{fig:Fig. 2}
 Left panel: The effective square lattice. Right panel:
 Phase diagram in the space spanned by $(\tilde J_x,\tilde J_y,J_z)$.
This is a (1,1,1) cross section with all J's positive.}
\end{figure}

\section{Mapping to a $p$-wave paired state}

The next step is to map the {\em Majorana fermion} $H_0$ to a
fermion model. Defining fermions on the $z$-links by \cite{cn}
$$d_s=(c_{s,b}+ic_{s,w})/2, ~~~d^\dag_s=(c_{s,b}-ic_{s,w})/2,$$
$H_0$ becomes
\begin{eqnarray}
H_0&=&J_z\sum_s(d_s^\dag d_s-1/2)+\tilde
J_x(d^\dag_sd_{s+e_x}-d_sd_{s+e_x}^\dag)\nonumber\\&+& \tilde
\lambda_x\sum_s(d^\dag_{s+e_x}d^\dag_s-d_{s+e_x}d_s)
\label{dh}\\&+& i\kappa_x\sum_s(d_sd_{s+e_x}+d_s^\dag
d_{s+e_x}^\dag)+y~{\rm partners}.\nonumber
\end{eqnarray}
This is a quadratic model of spinless fermions $d_s$ on the square
lattice (Fig.~\ref{fig:Fig. 2}, left panel) with general $p$-wave
pairing. If we include the 'z' partner of the three and four spin
terms in eq. (\ref{gk}), we have additional corresponding terms in
eq. (\ref{dh}) which are the next nearest neighbor terms in the
square lattice. These terms will not qualitatively affect our
result. Returning to Majorana fermions, a link fermion is a
superposition of two Majorana fermions in a link. Therefore, the
paring of the link fermions reflects the 'pairing' of Majorana
fermions.

After a Fourier transformation, eq. (\ref{dh}) becomes
\begin{eqnarray}
H_0&=&\sum_{\bf p} \xi_{\bf p}d^\dag_{\bf p}d_{\bf
p}+\frac{\Delta_{1,\bf p}}2(d_{\bf p}^\dag d^\dag_{-{\bf p}}+d_{\bf p} d_{-{\bf p}})\nonumber\\
&+&i\frac{\Delta_{2,\bf p}}2(d_{\bf p}^\dag d^\dag_{-{\bf
p}}-d_{\bf p} d_{-{\bf p}})
\end{eqnarray}
where the dispersion and the pairing functions are
\begin{eqnarray}
&&\xi_{\bf p}=J_z-\tilde J_x\cos p_x-\tilde J_y\cos
p_y,\nonumber\\
&&\Delta_{a,\bf p}=\Delta_{ax}\sin p_x+\Delta_{ay}\sin p_y, \quad
a=1,2\nonumber
\end{eqnarray}
with $\Delta_{1,x(y)}=\kappa_{x(y)}$ and
$\Delta_{2,x(y)}=\tilde\lambda_{x(y)}$. We have thus shown that
the ground state of the extended Kitaev model in Eq.~(1) are
equivalent to general $p$-wave paired fermion states. The
quasiparticle excitations are governed by the BdG equations
\begin{eqnarray}
E_{\bf p}u_{\bf p}=\xi_{\bf p}u_{\bf p}-\Delta^*_{\bf p}v_{\bf
p},~~E_{\bf p}v_{\bf p}=-\xi_{\bf p}v_{\bf p}-\Delta_{\bf p}u_{\bf
p}
\end{eqnarray}
where $E_{\bf p}=\sqrt{\xi_{\bf p}^2+(\Delta_{1,\bf
p})^2+(\Delta_{2,\bf p})^2}$ is the dispersion, $\Delta_{\bf
p}=\Delta_{1,{\bf p}}+i\Delta_{2,{\bf p}}$, and $(u_p,v_p)$ are
the coherence factors with $|u_{\bf
p}|^2=\frac{1}2(1+\frac{\xi_{\bf p}}{E_{\bf p}}),|v_{\bf
p}|^2=\frac{1}2(1-\frac{\xi_{\bf p}}{E_{\bf p}})$ and $v_{\bf
p}/u_{\bf p}=-(E_{\bf p}-\xi_{\bf p})/\Delta^*_{\bf p}$.

\section{Phase diagram in terms of the $p$-wave states}

We now turn to the properties of this $p$-wave paired state of
link fermions. It is instructive to consider the free fermion
dispersion. The condition $\xi_p=0$ defines a topological
transition between a band insulator and a metal with a Fermi
surface in the absence of pairing. The solution is given by $|\cos
p_x^*|=|\cos p_y^*|=1$, ${\bf p}^*=(0,0),(0,\pm\pi),(\pm
\pi,0),(\pm \pi,\pm \pi)$, where $J_z\pm \tilde J_x \pm \tilde
J_y=0$. Without loss of the generality, one considers only $\tilde
J_{x,y}>0$ and $J_z>0$. Then, $\xi_{p^*}=0$ corresponds to the
inner triangle of the (1,1,1)-cross section in $(\tilde J_x,\tilde
J_y,J_z)$ space (see Fig.~2 (right panel)). Notice that the p-wave
pairing gap functions $\Delta_{a,{\bf p}}$ vanish at ${\bf p}^*$
and therefore, this triangle is the gapless critical boundary
separating the A and B phases. Outside the triangle, $\xi_p>0$.
Thus the A phase is gapped. In the  limit ${\bf p}\to {\bf p}^*$,
the pair correlation $g_p\equiv v_p/u_p$ is analytic near ${\bf
p}^*$, implying tightly bound pairs in positional space and hence
the A-phase as the strong-pairing phase (See below for detailed
discussions) \cite{rg}. The global structure of the phase diagram
is invariant in the generalized $(\tilde J_x,\tilde J_y, J_z)$
space because our minimal three- and four-spin extension does not
change the topology of the underlying Fermi surface.

The nature of the B phase is much more intriguing. Inside the
triangle, $\xi_p$, $\Delta_{1,{\bf p}}$ and $\Delta_{2,{\bf p}}$ can
be zero individually. The gapless condition ($E_p=0$) requires all
three to be zero at a common ${\bf p}^*$. This can only be achieved
if (i) one of the $\Delta_{a,\bf p}=0$ or (ii) $\Delta_{1,\bf
p}\propto \Delta_{2,\bf p}$. If either (i) or (ii) is true,
$\xi_{\bf p}$ and $\Delta_{\bf p}$ can vanish simultaneously, i.e.
$E_p=0$ at ${\bf p}^*$, and the paired state is gapless. Otherwise,
the B phase is gapped. Note that contrary to conventional wisdom,
T-symmetry breaking alone does not guarantee a gap opening in the B
phase. The symmetry reason behind the gapless condition of the B
phase becomes clear in the continuum limit where $E_p=0$ implies
that the vortex-free Hamiltonian must be invariant, up to a
constant, under the transformation $p_x\leftrightarrow \eta p_y$ and
$\tilde J_x\leftrightarrow \eta^{-2} \tilde J_y$ with
$\eta=\frac{\Delta_{a,y}}{\Delta_{a,x}}$ with $a=1$ or $2$ and for
nonzero $\Delta$. We refer to this as a {\em generalized inversion
(G-inversion) symmetry} since it reduces to the usual mirror
reflection when $\eta=1$. This (projective) symmetry protects the
gapless nature of fermionic excitations and may be associated with
the underlying quantum order \cite{wz}. Kitaev's original model has
$\Delta_{1,i}=0$, and is thus G-inversion invariant and gapless. The
magnetic field perturbation \cite{ki2} breaks this G-inversion
symmetry and the fermionic excitation becomes gapped. A special case
with G-inversion symmetry breaking is $\Delta_{\bf p}\propto \sin
p_x+i\sin p_y$, i.e. the $p_x+ip_y$-wave paired state discussed by
Read and Green in the continuum limit \cite{rg}.

The $N$-fermion ground state wave function in the general $p$-wave
paired state can be written down as a Pfaffian for $N$ even: $
\Psi({\bf r}_1,...,{\bf r}_1)=$$1/(2^{N/2}(N/2)!)\sum_P {\rm sgn}
P\prod_{i=1}^{N/2}g({\bf r}_{P_{2i-1}}-{\bf r}_{P_{2i}}) $ where
$g(r)$ is the ``pair correlation'', i.e. the Fourier transform of
$g_{\bf p}=v_{\bf p}/u_{\bf p}$ in the BCS wave function $
|\Omega\rangle=\prod_{\bf p}|u_{\bf
p}|^{1/2}\exp{(\frac{1}2\sum_{\bf p} g_{\bf p}d^\dag_{\bf
p}d^\dag_{\bf -p})}|0\rangle$. The wave function exhibits very
different behaviors in the long wave length limit  in different
parameters \cite{note2}. In the A phase, $\xi_p>0$ as ${\bf
p}\to0$ , thus $g_{\bf p}\propto \Delta_{\bf p}$. The analyticity
of $g_{\bf p}$ leads to $g({\bf r})\propto e^{-\mu r}$ as in the
strong pairing phase of a pure $p_x+ip_y$ state \cite{rg}. In the
gapped B phase with G-inversion symmetry breaking, $\xi_p<0$ as
${\bf p}\to0$. Defining $ p_i'=\Delta_{ai}p_i$ with $a=1,2$ and
$i=x,y$, it follows that $g_{\bf p}\propto \frac{1}{p_x'+ip_y'}$,
leading to $g({\bf r})=\frac{1}{x_1'+ix_2'}$ with
$x'_a=\Delta^{-1}_{ai}x_i$ and thus a weak-pairing phase.
Identifying $z'=x'_1+ix_2'$, we see that the ground state of the
gapped B phase corresponds exactly to the Moore-Read Pfaffian. It
is easy to show that, $u$ and $v^*$ obey the same BdG equation in
this general $p$-wave paired state, such that the anti-particle of
the quasiparticle $\psi=(u,v)$ is itself, i.e.,  it is Majorana
fermion obeying Dirac equations in 2+1-dimensions \cite{rg}.

We now discuss the nature of the gapless B phase in the general
model with G-inversion symmetry. In this case, $E_{\bf p}=0$ at
${\bf p}=  \pm {\bf p}^*$ which are the solutions of $\xi_{\bf
p}=0$ and, say, $\Delta_{\bf p}=\Delta_{1,{\bf p}}=0$. At ${\bf
p}^*$, the fermion dispersions are generally given by 2D Dirac
cones. However, by a continuous variation of the parameters, one
can realize a dimensional reduction near the phase boundary where
the effective theory is in fact a 1+1 dimensional conformal field
theory in the long wave length limit. Let us consider parameters
that are close to the critical line with $|\sin p_a^*|\ll |\cos
p_a^*|$ where $g_{\bf q}={\rm sgn}[q_x\Delta_{1x}\cos
p_x^*+q_y\Delta_{1y}\cos p_y^*]\equiv{\rm sgn}(q_x')$ with ${\bf
q}={\bf p}-{\bf p}^*$. Doing the Fourier transform, we find
\begin{eqnarray}
g({\bf r})&=&\int dq'_xdq'_y
e^{iq_x'x'+iq_y'y}{\rm
sgn}(q_x')\nonumber\\
&=&\delta(y') \int dq_x' \frac{q_x'}{|q_x'|}\sin q_x'x'\sim
\frac{\delta(y')}{ x'}.
\end{eqnarray}
The $\delta(y)$-function indicates that the pairing in the gapless
B phase has a one-dimensional character and the ground state is a
one-dimensional Moore-Read Pfaffian. The BdG equations reduce to
\begin{eqnarray}
i\partial_t u=-i\Delta_{1x}(1+i\eta)\partial_{x'}v,~~i\partial_t
v=i\Delta_{1x}(1-i\eta)\partial_{x'}u,
\end{eqnarray}
with $\eta=\frac{\Delta_{1y}}{\Delta_{1x}}$. Thus, the gapless
Bogoliubov quasiparticles are one-dimensional Majorana fermions.
The long wave length effective theory for the gapless B phase near
the phase boundary is therefore the massless Majorana fermion
theory in 1+1-dimensional space-time, i.e. a $c=1/2$ conformal
field theory or equivalently a two-dimensional Ising model.

\section{Topological invariant in A and B phases}

We note that there is no spontaneous breaking of a continuous
symmetry associated with the phase transition from A to B phases.
Kitaev has shown that the A phase in his model is topologically
trivial and has zero spectral Chern number, while the gapped B phase
is characterized by the Chern number $\pm 1$ \cite{ki2}. This fact
was already discussed by Read and Green in the context of $p_x+ip_y$
paired state. Here we follow Read and Green \cite{rg} to study the
topological invariant in a general $p$-wave state.

In continuum limit, ${\bf p}=(p_x,p_y)$ lives in an Euclidean
space $R^2$. However, the constraint $|u_p|^2+|v_p|^2=1$
parameterizes a sphere $S^2$. As $|{\bf p}|\to\infty$, $\xi_{\bf
p}\to E_{\bf p}$ such that $v_{\bf p}\to 0$. Therefore, we can
compactify $R^2$ into an $S^2$ by adding $\infty$ to $R^2$ where
$v_{\bf p}\to 0$. The sphere $|u_{\bf p}|^2+|v_{\bf p}|^2=1$ can
also be parameterized by a pseudospin vector ${\bf
n_p}=(\Delta_{1,\bf p},-\Delta_{2,\bf p},\xi_{\bf p})/E_{\bf p}$
because $|{\bf n_p}|=1$. $(u_{\bf p},v_{\bf p})$ thus describes a
mapping from $S^2~ ({\bf p}\in R^2)$ to $S^2$ (spinor $|{\bf
n_p}|=1$). The winding number of the mapping is a topological
invariant.The north pole is $u_{\bf p}=1,v_{\bf p}=0$ at $|{\bf
p}|=\infty$ and the south pole is $u_{\bf p}=0,v_{\bf p}=1$ at
${\bf p=0}$. In the ${\bf n_p}$ parametrization, ${\bf
n_0}=(0,0,\frac{\xi_{\bf p}}{E_{\bf }})=(0,0,1)$ at $|{\bf
p}|=\infty$ and $(0,0,\frac{\xi_{\bf p}}{E_{\bf }})=(0,0,-1)$ at
${\bf p}=0$, corresponding to either the north pole or south pole.

In the strong pairing phase, we know that $u_p\to 1$ and $v_p\to 0$
as ${\bf p}\to 0$ (or equivalently, $\xi_{\bf p}>0$). This means
that for arbitrary ${\bf p}$, $(u_{\bf p},v_{\bf p})$ maps the
$p$-sphere to the upper hemisphere and the winding number is zero.
That is, the topological number $\nu=0$ in the strong pairing phase.

In the weak paring phase, $u_{\bf p}\to 0$ and $v_{\bf p}\to 1$ as
${\bf p}\to 0$. This means that the winding number is nonzero (at
least wrapping once). For our case, the winding number can be
directly calculated and is given by
\begin{eqnarray}
\nu&=&\frac{1}{4\pi}\int dp_xdp_y {\bf
n_p}\cdot\left(\partial_{p_x}{\bf n_p}\times
\partial_{p_y}{\bf n_p}\right)=1
\end{eqnarray}
Defining $P({\bf p})=\frac{1}2(1+{\bf n_p}\cdot {\vec \sigma})$,
which is the Fourier component of the projection operator to the
negative spectral space of the Hamiltonian, this winding number can
be identified as the spectral Chern number defined by Kitaev
\cite{ki2}
\begin{eqnarray}
\nu=\frac{1}{2\pi i} \int {\rm Tr}[P_-(\partial_{p_x}P_-
\partial_{p_y}P_--\partial_{p_y}P_-\partial_{p_x}P_-)]dp_xdp_y
\end{eqnarray}
where $P_-=I-P$ is a projective operator. This spectral Chern
number vanishes in the strong pairing A phase but takes an integer
value in the weak paring B phase. Thus, the phase transition from
A to B is a topological phase transition.

\section{vortex excitations}

We have discussed the universal behaviors of the ground state. We
now turn to discuss the $Z_2$ vortex excitation in the spin model
which corresponds to setting $W_P=-1$ for a given plaquette. The
Hamiltonian in the Majorana fermion representation is bilinear and
the energies of the vortices can be estimated both in the A phase
\cite{vidal} and the B phase\cite{pachos}. However, it remained
difficult to obtain analytical solutions of the wave functions
with two well-separated vortices. We have shown that the ground
state sector is equivalent to the $p_x+ip_y$ pairing theory for
fermions on the square lattice. Therefore, the Pfaffian state is
the ground state wave function in the continuum limit in the weak
pairing phase. Our strategy is to evaluate the energy of the trial
wave function containing vortices above the Pfaffian state in the
continuum limit. For two well separated half-vortices located at
$w_1$ and $w_2$ shown in Fig.~1 (lower panel), the Moore-Read
trial wave function has been well-studied \cite{mr,rg} and is
given by
\begin{eqnarray}
&&\Psi(z_1,...z_N;w_1,w_2)\propto {\rm
Pf}(g'(z_i,z_j;w_1,w_2)), \label{11}\\
&&g'(z_1,z_2;w_1,w_2)\propto\frac{(z_1-w_1)(z_2-w_2)+(w_1\leftrightarrow
w_2)}{z_1-z_2}.\nonumber
\end{eqnarray}
The second quantized state corresponding to this wave function
reads
\begin{eqnarray}
 |w_1,w_2\rangle\propto\exp\{\frac{1}2\sum_{{\bf r}_1,{\bf
r}_2}g'(z_1,z_2;w_1,w_2)d^\dag_{{\bf r}_1}d^\dag_{{\bf r}_2}\}
\end{eqnarray}
Performing a Fourier transformation, we have
\[
|w_1,w_2\rangle\propto\exp\{\frac{1}2\sum_{{\bf K},{\bf
k}}g'_k({\bf K})d_{\bf K+k}^\dag d_{\bf K-k}^\dag\},
\]
where ${\bf k}={\bf k}_1-{\bf k}_2$ and ${\bf K}={\bf k}_1+{\bf
k}_2$ are the relative and the total momenta of the pairs and
$g'_k({\bf K})$ is the Fourier transform of $g'({\bf r}_1,{\bf
r}_2)$. One can show that,
\begin{eqnarray}
&&g'_k({\bf K})\sim (\frac{1}k-\frac{A}{w_1w_2|{\bf
k}|^2k})\delta({\bf
K})\nonumber\\
&&+\frac{1}k(\frac{B}{w_1w_2|K|^2\bar
K^2}-\frac{(w_1+w_2)C}{w_1w_2|K|^2\bar K})\nonumber\\
&&=g_k'\delta({\bf K})+\frac{1}k \tilde g({\bf K})
\end{eqnarray}
where $A$, $B$ and $C$ are positive constants and $\tilde g({\bf
K})$ is independent of ${\bf k}$. Thus,
 \[|w_1,w_2\rangle\propto \exp
 \{\frac{1}2\sum_{\bf k}g'_kd_{\bf k}^\dag d_{\bf -k}^\dag+\sum_{\bf K,k}
 (1/k)g'({\bf K})
 d_{\bf K+k}^\dag d_{\bf K-k}^\dag\}
 \]
Such a vortex pair is shown in Fig.~1 where the red $z$-links have
$u_{bw}=-1$ and all others have $u_{bw}=1$. The corresponding
Hamiltonian can be written as $H=H_0+\delta H$, where $H_0$ is the
vortex-free Hamiltonian and $\delta H$ is the vortex part. The
latter is expressed as a sum of the pairing and chemical potential
terms according to Eq.~(5) over the red $z$-links extending in the
$\xi$-direction (i.e., the direction with $x=y$) between the
vortices. It is straightforward to show that $\delta H$ has the
following expectation value in the vortex state,
\begin{eqnarray}
&&\langle w_1,w_2|\delta H|w_1,w_2\rangle\\
&&\propto
\sum_{p_\xi,p'_\xi}\frac{i(e^{iw_1(p_\xi+p'_\xi)}-e^{iw_2(p_\xi+p'_\xi)})}
{p_\xi+p'_\xi}f(p_\xi,p_\xi')=0,\nonumber
\end{eqnarray}
where $f(p_\xi,p_\xi')$ is an analytical function of
$p_\xi+p'_\xi$. \emph{This means that there are no a continuum
spectrum above the vortex pairs and then the vortex pairs are also
separated from other higher energy excitations.} On the other
hand, one can check that since $[H_0,\sum_{\bf K,k}g'_k({\bf K}\ne
0)d_{\bf K+k}^\dag d_{\bf K-k}^\dag]=0$, the ${\bf K}\ne0$ sector
does not play a nontrivial role in calculating the energy $E_v$ of
such a vortex pair. The latter is given by
\begin{eqnarray}
&&E_v=\langle w_1,w_2|H|w_1,w_2\rangle=\langle w_1,w_2|H_0|w_1,w_2
\rangle\nonumber\\
&&=\sum_{\bf k}E_{\bf k}|u_{\bf k}\delta g_{\bf k}|^2\langle
w_1w_2|d_{\bf k}d^\dag_{\bf k}|w_1w_2\rangle\nonumber\\
&&= \sum_{\bf k} E_{\bf k}|u_{\bf k}\delta g_{\bf
k}|^2/(1+|g'^0_k|^2),
\end{eqnarray}
where $g'^0_k=g'_k+\frac{1}k(\frac{B}{w_1w_2\bar
K^2}-\frac{(w_1+w_2)C}{w_1w_2\bar K})|_{{\bf K}\to 0}$ and $\delta
g_{\bf k}=g'^0_k-g_{\bf k}$. Physically, the factor $\langle
w_1w_2|d_{\bf k}d^\dag_{\bf
k}|w_1w_2\rangle=1-|g'^0_k|^2/(1+|g'^0_k|^2)=1/(1+|g'^0_k|^2)$ in
Eq.(15) is the quasihole distribution when the two vortices are
located at $w_1$ and $w_2$. Thus, $E_v$ indeed corresponds to the
energy cost to excite the vortex pair. We have evaluated the
vortex pair energy $E_v$ in different limits. First, if $w_1$ and
$w_2$ were sent to infinity before ${\bf K}\to 0$, then $g'^0_k\to
g_k$ and we recover the ground state. Second, if ${\bf K}\to 0$
while $w_1$ and $w_2$ remain finite, then $E_{\bf k}|u_{\bf
k}\delta g_{\bf k}|^2/(1+|g'^0_k|^2)\sim E_{\bf k}|u_{\bf k}|^2$,
which tends to $|{\bf k}|^2$ in both the small and large $k$
limits. The excitation energy is thus high but finite due to the
short distance cut-off. (Recall that $\xi_{\bf k}<0$ in the B
phase). It is independent of $w_1$ and $w_2$ and as a result the
vortices are deconfined. The third case is when the vortices are
far from the origin such that $|Kw_{1}|$ and $|Kw_2|$ are finite.
In the short distance, large $k$ limit, $E_{\bf k}|u_{\bf k}\delta
g_{\bf k}|^2/(1+|g'^0_k|^2)\sim |{\bf k}|^{-4}$. On the other
hand, in the long wavelength limit with $|{\bf k}|\to 0$, $|u_{\bf
k}|\sim |{\bf k}|$, $|g'^0_k|\to |{\bf k}|^{-3}$ and $E_{\bf
k}\to$ constant such that $E_{\bf k}|u_{\bf k}\delta g_{\bf
k}|^2/(1+|g'^0_k|^2)\sim |{\bf k}|^2\to 0$. As a result, the
vortex pair energy $E_v$ is free of infrared divergences and is
only weakly dependent on $w_1-w_2$. Therefore, the vortices are
also deconfined. Finally, since $E_k$ increases with the paring
parameters $\kappa$ and $\lambda$ in Eq.(4), $E_v$ is expected to
increase with increasing paring gap parameters. These results are
consistent with the finite size numerical calculations of the
vortex pair energy in the lattice model \cite{pachos}. Our
analytical results suggest that the vortex pair described by
Eq.(11), while costing a high energy in the bulk, corresponds to
low energy excitations near the edge of the system.

We note an important difference between this $p$-wave theory and a
conventional $p$-wave superfluid: Instead of spontaneously breaking
the $U(1)$ symmetry in an usual $p$-wave superfluid, only the
discrete $Z_2$ symmetry is broken and the $U(1)$ symmetry is absent
in the present model. The vortices studied here are thus $Z_2$
vortices instead of $U(1)$ vortices. As a consequence, in the gapped
B phase, the vortices are in the deconfinement phase \cite{sen}
instead of being logarithmically confined in the $p$-wave
superfluid. This fact can be easily seen because $\Delta_{ax(y)}$
are real and there is no $U(1)$ phase factor whose gradient gives
rise to a vector field of the vortex. Our analysis of the vortex
pair energy also shows this difference.

The finiteness of $E_v$ and the vanishing of $\langle \delta
H\rangle=0$ imply that the vortex excitations are {\it separated}
either from the ground state or other excitations. This is
consistent with analysis of Read and Green on the U(1) vortex
excitations in the $p$-wave paired state \cite{rg,dhlee}. To
determine how close the Moore-Read vortex state is to the exact
vortex excitations in this model requires a numerical calculation
of the overlapping between the exact eigenstates and the
Moore-Read vortex wave functions. A more important question is the
realization of the Read-Moore four-vortex state which has a
two-fold degeneracy with the vortices obeying non-abelian
statistics \cite{mr}. We leave these studies to future works.

\section{conclusions}

We have constructed an exactly soluble spin model with two-, three-
and four-spin couplings on a honeycomb lattice. The ground state
sector of this model on the honeycomb lattice is mapped to a
$p$-wave paired state of the link fermions on a square lattice with
general pairing parameters. Based on the general $p$-wave paired
states, we analyzed the phase diagram of the system and the
properties of topologically different phases. We found that our
phase diagram is universal and includes both the Kitaev model and
the Pfaffian state in its universality class.

The authors thank the KITPC for warm hospitality where this work
was initiated and finalized during the Program on Quantum Phases
of Matter. The authors are grateful to Z. Nussinov, N. Read, X.
Wan, X.-G. Wen, K. Yang, J. W. Ye, and Y. S. Wu for useful
discussions. This work was supported in part by NNSF of China, the
Ministry of Science and Technology of China, a fund from CAS and
the U.S. DOE Grant No. DE-FG02-99ER45747.


\vspace{-0.5cm}

\begin{references}

\vspace{-0.5cm}


\bibitem{mr} G. Moore and N. Read, Nucl. Phys. B {\bf 360}, 362 (1991).
\bibitem{xia} J. S. Xia, W. Pan, C. L. Vincente, E. D. Adams, N. S. Sullivan,
H. L. Stormer, D. C. Tsui, L. N. Pfeiffer, K. W. Baldwin, K. W.
West, Phys. Rev. Lett. {\bf 93}, 176809 (2004).
\bibitem{ki1} A. Kitaev, Ann. Phys. {\bf 303}, 2(2003). M. H. Freedman,
 M. Larsen, Z. H. Wang, Commun. Math. Phys. {\bf 227}, 605(2002).


\bibitem{ki2} A. Kitaev, Ann. Phys. {\bf 321}, 2(2006).

\bibitem{fzx} X. Y. Feng, G. M. Zhang, and T. Xiang, Phys. Rev. Lett. {\bf 98}, 087204 (2007).

\bibitem{GWW} M. Greiter, X.G. Wen and F. Wilczek, Nucl. Phys. B {\bf 374}, 567
(1992).

\bibitem{rg} N. Read and  D. Green, Phys. Rev. B {\bf 61}, 10267 (2000).

\bibitem{lieb} E. H. Lieb,  Phys. Rev. Lett. {\bf 73}, 2158 (1994).

\bibitem{cn} H. D. Chen and J. P. Hu,
Phys. Rev. B {\bf 76}, 193101 (2007).
 H. D. Chen and Z. Nussinov, J. Phys. A {\bf 41}, 075001 (2008).


\bibitem{vidal} K. P. Schmidt, S. Dusuel and J. Vidal,
Phys. Rev. Lett. {\bf 100}, 057208 (2008).

\bibitem{pachos}  V. Lahtinen, G. Kells, A. Carollo, T.
Stitt, J. Vala and J. K. Pachos,  Ann. of Phys. {\bf 323}, 2286
(2008).

\bibitem{wz} X. G. Wen and A. Zee, Phys.Rev. B {\bf 66}, 235110
(2002).


\bibitem{note2} There are other gapless lines corresponding to ${\bf
p}^*=(0,\pi)$ and $(\pi,0)$. The Pfaffian and anti-Pfaffian states
may be discussed near these points. For the anti-Pfaffian, see, S.
S. Lee, S. Ryu, C. Nayak, and M. P. A. Fisher, Phys. Rev. Lett.
{\bf 99}, 236807 (2007) ; M. Levin, B. Halperin, and B. Roscnow,
Phys. Rev. Lett. {\bf 99}, 236806 (2007).



\bibitem{dhlee} D. H. Lee, G. M. Zhang, and T. Xiang, Phys. Rev. Lett. {\bf 99}, 196805 (2007).

\bibitem{sen} T. Senthil and M. P. Fisher, Phys. Rev. B {\bf 62}, 7850
(2000).


\end{references}
\end{document}